\pdfoutput=1
\documentclass[sigconf,authorversion,nonacm]{acmart}

\AtBeginDocument{%
  }

\usepackage{subcaption}

\usepackage{enumitem}

\newtheorem{problem}{Problem}

\begin{document}

\title{ReCon: Reducing Congestion in Job Recommendation using Optimal Transport}

\author{Yoosof Mashayekhi}
\email{yoosof.mashayekhi@ugent.be}
\orcid{0000-0002-8993-0751}
\affiliation{%
  \institution{IDLAB - Department of Electronics and Information Systems, Ghent University}
  \city{Ghent}
  \country{Belgium}
  \postcode{9000}
}

\author{Bo Kang}
\email{bo.kang@ugent.be}
\orcid{0000-0002-9895-9927}
\affiliation{%
  \institution{IDLAB - Department of Electronics and Information Systems, Ghent University}  \city{Gent}
  \country{Belgium}
  \postcode{9000}
}

\author{Jefrey Lijffijt}
\email{jefrey.lijffijt@ugent.be}
\orcid{0000-0002-2930-5057}
\affiliation{%
  \institution{IDLAB - Department of Electronics and Information Systems, Ghent University}  \city{Gent}
  \country{Belgium}
  \postcode{9000}
}

\author{Tijl De Bie}
\email{tijl.debie@ugent.be}
\orcid{0000-0002-2692-7504}
\affiliation{%
  \institution{IDLAB - Department of Electronics and Information Systems, Ghent University}  \city{Gent}
  \country{Belgium}
  \postcode{9000}
}
\renewcommand{\shortauthors}{Mashayekhi et al.}

\begin{abstract}
Recommender systems may suffer from \emph{congestion}, meaning that there is an unequal distribution of the items in how often they are recommended. Some items may be recommended much more than others. Recommenders are increasingly used in domains where items have limited availability, such as the job market, where congestion is especially problematic: Recommending a vacancy---for which typically only one person will be hired---to a large number of job seekers may lead to frustration for job seekers, as they may be applying for jobs where they are not hired. This may also leave vacancies unfilled and result in job market inefficiency.

We propose a novel approach to job recommendation called \texttt{ReCon}, accounting for the congestion problem. Our approach is to use an optimal transport component to ensure a more equal spread of vacancies over job seekers, combined with a job recommendation model in a multi-objective optimization problem. We evaluated our approach on two real-world job market datasets. The evaluation results show that \texttt{ReCon} has good performance on both congestion-related (e.g., Congestion) and desirability (e.g., NDCG) measures.\end{abstract}




\maketitle
\section{Introduction}\label{sec:intro}
Recommendation systems are widely used in online platforms \cite{batmaz2019review}. They are designed to enhance the user experience by providing personalized recommendations \cite{schafer2001commerce,guo2020survey} to facilitate access to the most relevant content. Recommendation systems have been applied to various types of content, such as movies \cite{DBLP:conf/aaai/WangZZQX20,zhuang2017representation}, online shopping \cite{han2019adaptive,xu2019neuo}, and jobs \cite{qin2020enhanced}.

\begin{figure}[t]
  \centering
  \includegraphics[width=1\linewidth,height=\textheight,keepaspectratio]{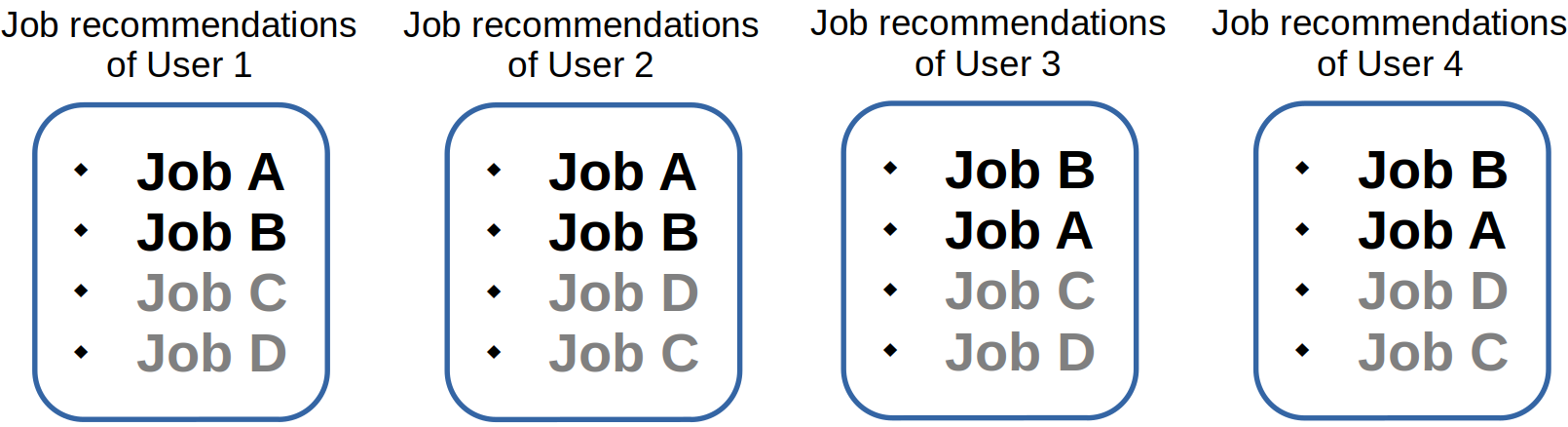}
  \caption{An illustration of congestion in job recommendation. Four jobs are recommended to four users, while the top-2 recommendations only include jobs A and B for all users (e.g., because of the popularity bias or their inherent higher relevancy score). Hence, if users only apply for their top-2 recommendations, jobs A and B receive four applications, while jobs C and D receive none.}\label{fig:congestion_illustration}
  \Description[Illustration of the congestion problem]{An example of congestion in job recommendation.}
\end{figure}

The aim of job recommendation is to recommend jobs (items) that are suitable for job seekers (users) and, if applied correctly, can have a positive impact on their career paths. 
In a job market, there is competition between job seekers for jobs and also between jobs (or companies) for job seekers \cite{mashayekhi2022challenge}. This competition may be aggravated by the presence of so-called \emph{congestion}, meaning that some items are more visible and desirable than others \cite{chen2019prediction}. The congestion problem is illustrated with an example in Figure~\ref{fig:congestion_illustration}. The use of job recommender systems risks to create or amplify congestion, as desirable items are often recommended to many users. This risks frustrating job seekers, as they will be unlikely to be successful when all applying for the same jobs. It is therefore important that job recommender systems explicitly account for congestion and try to limit or reduce it. The aim of this paper is to address that need.

\begin{problem}\label{problem}
The problem we address in this paper is to reduce congestion in job recommendation, i.e., to improve congestion-related measures such as Congestion \cite{bied2021congestion}, Coverage, and Gini index, while recommending desirable recommendations, i.e., to keep good performance of desirability measures such as Recall, NDCG, and Hit Rate.
\end{problem}

Our approach is to spread jobs over job seekers more equally using the optimal transport theory. We optimize the spread of jobs over job seekers with a given job recommendation model as a multi-objective task. Since optimal transport aims to match two distributions, we minimize the optimal transport between job seekers and jobs, to enforce a high matching degree between each job and a few job seekers and vice versa. Hence, we expect that the congestion in the generated recommendations by the model is reduced. 

The joint optimization of the base recommendation model and the optimal transport component leads to having a trained model that offers advantages over post-processing approaches in real-life scenarios that require incremental updates of the model.


The main \textbf{contributions} of this paper are:
\begin{itemize}
\item We introduce a novel approach, \texttt{ReCon}, to reduce congestion in a given job recommendation model. We jointly optimize the job recommendation model objective function and optimal transport cost between job seekers and jobs as a multi-objective task. (Section~\ref{sec:proposed_method})
\item We show how to employ the so-called entropic optimal transport using the Sinkhorn algorithm \cite{cuturi2013sinkhorn} to facilitate the optimization in \texttt{ReCon}. (Section~\ref{sub_sec:optimization})
\item We evaluate \texttt{ReCon} on two job recommendation datasets and compare it with the baselines in terms of three desirability measures (e.g., NDCG), and also three congestion-related measures (e.g., Congestion). (Section~\ref{sec:experimental_evaluation})
\end{itemize}
Related work is summarized in Section~\ref{sec:related_work} and final conclusions are presented in Section~\ref{sec:conclusion_future_work}.

\section{Proposed Method}\label{sec:proposed_method}
In this section, we introduce our approach, \texttt{ReCon}, to address Problem~\ref{problem}, reducing congestion in job recommendation. We first give a brief introduction to optimal transport theory and its efficient optimization using the Sinkhorn algorithm in Section~\ref{sub_sec:background_ot}. Next, we explain the multi-objective approach in Section~\ref{sub_sec:ReCon}. We then introduce the requirements for the efficient optimization of \texttt{ReCon} in Section~\ref{sub_sec:optimization}. Finally, we suggest a matching cost function and a similarity function for computing optimal transport between job seekers and jobs in Section~\ref{sub_sec:proposed_method_matching_cost} that meet those requirements.

\subsection{Optimal Transport}\label{sub_sec:background_ot}

Optimal transport refers to the problem of finding the most efficient way to transport one given distribution of resources (e.g., masses, probabilities) to another. Mathematically, this can be represented as finding a transportation plan that minimizes the cost of transporting the resources from one place to another, subject to certain constraints. The cost of transporting the resources is often represented using a distance function. In the context of the present paper, we are interested in the optimal transport between users  $U$ (job seekers) and items $I$ (jobs), where $\boldsymbol{D}=[d_{ui}]$ denotes a cost of matching users with items. Hence, the optimal transport plan $\boldsymbol{F}=[f_{ui}]$ is given as:
\begin{align}
\min_{\boldsymbol{F}} \quad &\sum_{u \in U}\sum_{i \in I} f_{ui}d_{ui}, \quad
\textrm{s.t.} \quad f_{ui}  \geq  0 \quad \forall (u,i) \in U \times I, \nonumber \\
&\sum_{i \in I}f_{ui} = w_u \quad \forall u \in U, \quad
\sum_{u \in U}f_{ui} = w_i \quad \forall i \in I. \label{eq:ot_exact_eq}
\end{align}

A tractable relaxation of the above optimization problem is proposed in \cite{cuturi2013sinkhorn}, by
regularization with an entropy term:
\begin{align}
\min_{\boldsymbol{F}} \quad &\sum_{u \in U}\sum_{i \in I} f_{ui}(d_{ui}+\epsilon\log(f_{ui})), \quad
\textrm{s.t.} \quad \sum_{i \in I}f_{ui} = w_u \quad \forall u \in U, \nonumber \\
&\sum_{u \in U}f_{ui} = w_i \quad \forall i \in I, \label{eq:ot_sinkhorn_eq}
\end{align}

with $\epsilon$ the regularization weight. This relaxed optimization problem can be solved with the Sinkhorn algorithm \cite{cuturi2013sinkhorn}, where its solution is differentiable w.r.t. the cost values $\boldsymbol{D}$, and suitable for various optimization tasks. 

For simplicity, we set $w_u=\frac{1}{\vert U \vert}$ for all users $u \in U$ and $w_i=\frac{1}{\vert I \vert}$ for all items $i \in I$ in Eq.~(\ref{eq:ot_exact_eq}) and Eq.~(\ref{eq:ot_sinkhorn_eq}), so that the total mass for users and items is the same and all users, respectively items, are treated equally.

\subsection{ReCon}\label{sub_sec:ReCon}
Our approach is to optimize the desirability objective function of a recommendation model and an optimal transport cost between job seekers and jobs as a multi-objective task. The costs of matching individuals to jobs in the optimal transport are based on the recommendation scores, such that it uses the preferences of each individual, but it goes beyond the objective of the recommender, as the optimal transport problem accounts also for the preferences of other individuals. The optimal transport problem distributes job seekers more equally over all jobs and vice versa.
Hence, we expect that the recommendations from \texttt{ReCon} have lower congestion.

For a given recommendation model M with an objective function $O_{\text{M}}$ and a set of parameters $\boldsymbol{P}=[p_{ui}]$, where $p_{ui}$ represents the matching score between user $u$ and item $i$, the multi-objective task is as follows:
\begin{align}
O_{\text{ReCon}} = O_{\text{M}} + \lambda O_{\text{C}}, \label{eq:multi-O_eq-v1}
\end{align}
where $O_{\text{C}}$ is the objective function of the optimal transport between job seekers and jobs that is defined using the recommendation scores $\boldsymbol{P}$ (defined in Section~\ref{sub_sec:proposed_method_matching_cost}). The goal of $O_{\text{C}}$ is to reduce the congestion in the recommendations. The weight of $O_{\text{C}}$ in the multi-objective task is also indicated by the hyper-parameter $\lambda$.

To recommend top-$k$ items to user $u$, we have to compare the recommendation scores between $u$ and all items. To account for this competition between different items for the same user, it is important not only to penalize the cost of matching user-item pairs that are part of the optimal transport solution in the optimal transport objective function, but also to penalize the similarity of user-item pairs that are not part of the optimal transport solution (more intuition about this approach is provided in Section~\ref{sub_sec:proposed_method_matching_cost}). Hence, the optimal transport optimization problem in \texttt{ReCon} is:
\begin{align}
\min_{\boldsymbol{F}} \quad &\sum_{u \in U}\sum_{i \in I} f_{ui}c(p_{ui}) + (1-f_{ui})s(p_{ui}), \nonumber \\
&\textrm{s.t.} \quad f_{ui}  \geq  0 \quad \forall (u,i) \in U \times I, \nonumber \\ 
&\sum_{i \in I}f_{ui} = w_u \quad \forall u \in U, \quad
\sum_{u \in U}f_{ui} = w_i \quad \forall i \in I, \label{eq:ot_modified_v1}
\end{align}

where $c$ is a function of $\boldsymbol{P}$ to indicate the cost of matching a user and an item and $s$ is a function of $\boldsymbol{P}$ to indicate the similarity of a user and an item. We can write the objective function in Eq.~(\ref{eq:ot_modified_v1}) as $\sum\limits_{u \in U}\sum\limits_{i \in I} s(p_{ui}) + \sum\limits_{u \in U}\sum\limits_{i \in I}f_{ui}(c(p_{ui}) - s(p_{ui}))$. Since the first double summation does not depend on $\boldsymbol{F}$, it can be optimized outside the linear program in Eq.~(\ref{eq:ot_modified_v1}). Thus, the objective function in the linear program can be written as $\sum\limits_{u \in U}\sum\limits_{i \in I}f_{ui}(c(p_{ui}) - s(p_{ui}))$, which can be optimized efficiently using the Sinkhorn algorithm with Eq.~(\ref{eq:ot_sinkhorn_eq}), using $d_{ui}=c(p_{ui}) - s(p_{ui})$. We denote this objective function by $O_{\text{COT}}$. Hence, the multi-objective task can be written as:
\begin{align}
O_{\text{ReCon}} = O_{\text{M}} + \lambda O_{\text{COT}} + \lambda \sum_{u \in U}\sum_{i \in I} s(p_{ui}). \label{eq:multi-O_eq-v2}
\end{align}

 

\subsection{Efficient optimization}\label{sub_sec:optimization}



For efficient optimization, all terms in Eq.~(\ref{eq:multi-O_eq-v2}) have to be differentiable w.r.t. $\boldsymbol{P}$. Since $O_{\text{COT}}$ computed by Eq.~(\ref{eq:ot_exact_eq}) is not differentiable w.r.t. functions $c$ and $s$, it may also not be differentiable w.r.t. $\boldsymbol{P}$ because the functions $c$ and $s$ depend on the recommendation scores $\boldsymbol{P}$. If we compute $O_{\text{COT}}$ by Eq.~(\ref{eq:ot_sinkhorn_eq}), the Sinkhorn algorithm could be used for the optimization, which is differentiable w.r.t. $\boldsymbol{P}$ when $c$ and $s$ are differentiable w.r.t. $\boldsymbol{P}$. We also require the objective function of the recommendation model $O_{\text{M}}$ to be differentiable w.r.t. $\boldsymbol{P}$. Hence, the objective function $O_{\text{ReCon}}$ in Eq.~(\ref{eq:multi-O_eq-v2}) would be differentiable w.r.t. $\boldsymbol{P}$ and optimization algorithms such as gradient descent may be employed for the optimization.



\textbf{In summary, the requirements} for efficient optimization are:
\begin{enumerate}[topsep=2pt]
\item The objective function of the recommendation model $O_{\text{M}}$ must be differentiable w.r.t. $\boldsymbol{P}$.
\item Functions $c$ and $s$ must be differentiable w.r.t. $\boldsymbol{P}$.
\end{enumerate}
We now propose a matching cost function $c$ and a similarity function $s$ that meet the second requirement.


\subsection{The matching cost and similarity functions}\label{sub_sec:proposed_method_matching_cost}
High recommendation score $p_{ui}$ indicates that user $u$ and item $i$ may match well. Hence, the cost of matching them in optimal transport should be low and their similarity should be high. If $p_{ui}>0$, we suggest using the following functions $c$ and $s$ as the cost of matching user $u$ with item $i$ and the similarity of user $u$ and item $i$ respectively:
\begin{align}\label{eq:matching_cost}
 c(p_{ui}) = -\ln(p_{ui}), \quad s(p_{ui}) = -\ln(1-p_{ui}).
\end{align}
%
%
Hence, $O_{\text{C}}$ in Eq.~(\ref{eq:multi-O_eq-v1}) would be $-\sum\limits_{u \in U}\sum\limits_{i \in I} f_{ui}\ln(p_{ui}) + (1-f_{ui})ln(1-p_{ui}) = -\sum\limits_{u \in U}\sum\limits_{i \in I} \ln(p_{ui}^{f_{ij}})+\ln((1-p_{ui})^{1-f_{ij}}) = -\ln\prod\limits_{u \in U}\prod\limits_{i \in I} p_{ui}^{f_{ui}} (1-p_{ui})^{1-f_{ui}}$.

The intuition behind choosing this term is as follows. In the case where $|U| = |I|$ the optimal transport has an optimum where each user is mapped to exactly one item and vice versa, i.e., $f_{ij} \in \{0,1\}$ in $\boldsymbol{F}=[f_{ij}]$. For this case, if $\boldsymbol{P}$ represents the matching probabilities, it is easy to see that computing the optimal transport would be the same as computing a matching between users and items with the highest probability accounting for the user-item pairs that are not part of the matching as well.

Functions $c$ and $s$ in Eq.~(\ref{eq:matching_cost})
 are differentiable w.r.t. the recommendation scores $\boldsymbol{P}$ (the set of parameters of model M). Hence, functions $c$ and $s$ in Eq.~(\ref{eq:matching_cost})
   meet the second requirement in Section~\ref{sub_sec:optimization}, allowing for efficient optimization.





\section{Experimental Evaluation}\label{sec:experimental_evaluation}

In this section, we describe the experimental evaluation of \texttt{ReCon}. We investigate two research questions \textbf{Q1}: How does \texttt{ReCon} perform in congestion-related measures and desirability measures compared to the base recommendation model and the baselines? \textbf{Q2}: How does \texttt{ReCon} perform in terms of execution time compared to other methods? 

We first discuss the datasets, recommendation model, baselines, and settings. Next, we present the result of each experiment. The source code for repeating the experiments, the preprocessed public datasets, details of the selection of hyper-parameters for each dataset, more results for baseline comparison, and hyper-parameter sensitivity analysis for $\lambda$ are available in the  \href{https://github.com/aida-ugent/ReCon/}{online supplementary material}\footnote{\url{https://github.com/aida-ugent/ReCon/}}.




\subsection{Datasets}\label{sub_sec:experimental_evaluation_datasets}
We evaluated the methods using two datasets:

\textbf{VDAB}\footnote{\url{https://www.vdab.be/}}: VDAB is the employment service of Flanders in Belgium. 
The dataset contains a suitably anonymized sample of the applications made by job seekers to available job vacancies in the last ten days of 2018.

\textbf{CareerBuilder}\footnote{\url{https://www.careerbuilder.com/}}: CareerBuilder is an e-recruitment platform, which matches job seekers with jobs. We use the applications made by job seekers to available job vacancies in the last ten days of its public dataset\footnote{\url{https://www.kaggle.com/c/job-recommendation}} for the evaluation.

To better use collaborative filtering signals, we only keep job seekers and jobs with at least four interactions in both datasets. Table~\ref{tab:datasets} shows the main statistics of the datasets. 

\begin{table*}[]
\centering
  \caption{Main statistics of the datasets used for evaluation.}
  \label{tab:datasets}
  \setlength\tabcolsep{4pt}
  \begin{tabular}{lccccc}
    \toprule
    Dataset & Job seekers & Jobs & Train Interactions & Validation Interactions & Test Interactions \\
    \midrule
    VDAB & 1693 & 2931 & 9766 & 1428 & 2950 \\
    CareerBuilder & 3876 & 4337 & 24316 & 1071 & 4557 \\
    \bottomrule
  \end{tabular}
\end{table*}

\subsection{Recommendation model}\label{sub_sec:experimental_evaluation_recommendation_model}
Although \texttt{ReCon} can be applied to any recommendation model satisfying the requirements in Section~\ref{sub_sec:optimization}, we used a collaborative filtering method for the evaluation. The reason for choosing a CF method is that the performance of content-based methods depends on the quality of the contextual features whereas obtaining those features requires different feature engineering pipelines for different datasets. Since the focus of this paper is not improving the recommendation performance with different feature engineering techniques, we evaluated \texttt{ReCon} using a CF method.

We use Conditional Network Embedding (CNE; \cite{kang2018conditional}) as the base recommendation model. CNE is a network embedding method that proposes a probability distribution for the network conditional on the embedding and finds the optimal embedding by maximum likelihood estimation. CNE satisfies the requirements in Section~\ref{sub_sec:optimization} for efficient optimization.

\subsection{Baselines}\label{sub_sec:experimental_evaluation_baselines}
We compare \texttt{ReCon} with the following baselines that both receive the recommendation scores from a trained model in their first step. Hence, both baselines have post-processing approaches.

\textbf{CAROT} \cite{bied2021congestion}: CAROT employs optimal transport to redistribute jobs among job seekers. The final recommendation is according to the optimal transport solution.
Since both methods presented in \cite{bied2021congestion} and \cite{naya2021designing} have a similar approach, we only compare our proposed method with CAROT \cite{bied2021congestion}.

\textbf{FairRec} \cite{patro2020fairrec}: FairRec is a greedy method that ensures Envy-Free job recommendations for job seekers and guarantees a minimum exposure for job vacancies by converting the fair recommendation problem into a fair allocation problem. Since \cite{patro2020fairrec} and \cite{biswas2021toward} have a similar approach and the modification in \cite{biswas2021toward} is not focused on the congestion problem, we only compare our proposed method with FairRec \cite{patro2020fairrec}. As FairRec requires $|I| \leq k \cdot |U|$, we did not evaluate it for $k = 1$.

\subsection{Settings}\label{sub_sec:experimental_evaluation_settings}
Both datasets include the interactions in a period of ten days. For each dataset, we use the interactions in the first six days for training, the seventh day for validation, and the last three days for the test set. We also use the source code provided by the authors for the baselines. 

For Carot, we select $\epsilon$ from \{1, 10, 100\}. We also use four different functions to obtain the matching costs in the second step of the method (that solves the optimal transport problem) including linear (Id+), exponential (Exp+), rank-based (Rank), or NDCG-like (NDCG) functions as mentioned in CAROT paper \cite{bied2021congestion}.
For FairRec, we select $\alpha$ (the coefficient for the producer-side guarantee in their method) from \{0.2, 0.4, 0.6, 0.8, 1.0\}. Since FairRec generates a specific list of $k$ recommendations for a given $k$ that is not guaranteed to contain the same order of recommendations for e.g., $k-1$, we run the method for each value $k$ in the top-$k$ recommendation for evaluation.
For \texttt{ReCon}, we set $\epsilon$ to 10, the number of Sinkhorn iterations to 10, and we select $\lambda$ from \{1e-1, 1e-2, 1e-3, 1e-4, 1e-5, 1e-6\}. 

We evaluate the methods on the following desirability measures:

\begin{itemize}
\item NDCG: Normalized discounted cumulative gain, which is computed by summing the true scores ranked in the order induced by the predicted scores after applying a logarithmic discount, divided by the best possible score.

\item Recall: The proportion of relevant items found in the top-$k$ recommendations.

\item Hit Rate: The fraction of users for which the correct answer is included in the top-$k$ recommendations.
\end{itemize}

We also evaluate the methods on the following congestion-related measures:

\begin{itemize}
\item Congestion: Congestion is defined as the negative entropy of item market shares, where item market share $MS_l(i)$ of item $i$ is defined as the fraction of users such that item $i$ appears among their top-$k$ recommendations \cite{bied2021congestion}. As suggested by Bied et al. \cite{bied2021congestion}, Congestion is divided by $\log(\vert I \vert)$ ($I$ represents the set of items) to map it to $[-1, 0]$. Congestion is minimized if the entropy of the market shares is minimized. Hence, -1 is the optimal value. 

\item Coverage: The fraction of items involved in top-$k$ recommendation of at least one user \cite{bied2021congestion}.

\item Gini Index: The Gini index is a statistical measure of the inequality of a distribution \cite{piketty2015economics}. In the context of job recommendation, the Gini index is computed for the market share of items.
\end{itemize}

\subsection{Results}\label{sub_sec:experimental_evaluation_results}
In this section, we present the results of the experiments to investigate the research questions \textbf{Q1} and \textbf{Q2}.

\subsubsection{Baseline comparison\label{exp:baseline_comparison}}
\begin{figure*}[]
  \centering
  \begin{subfigure}{0.6\textwidth}
  \includegraphics[width=\textwidth,height=\textheight,keepaspectratio]{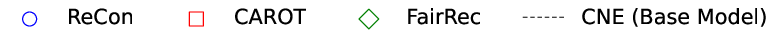}

  \end{subfigure}

  \begin{subfigure}{1\textwidth}
  \begin{subfigure}{.33\textwidth}
  \includegraphics[width=\textwidth,height=\textheight,keepaspectratio]{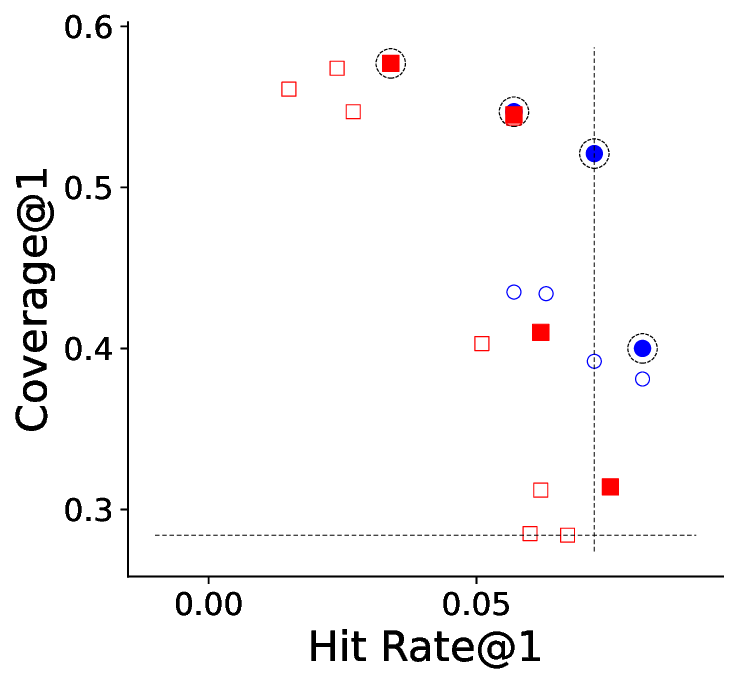}
  \end{subfigure}
  \begin{subfigure}{.33\textwidth}
  \includegraphics[width=\textwidth,height=\textheight,keepaspectratio]{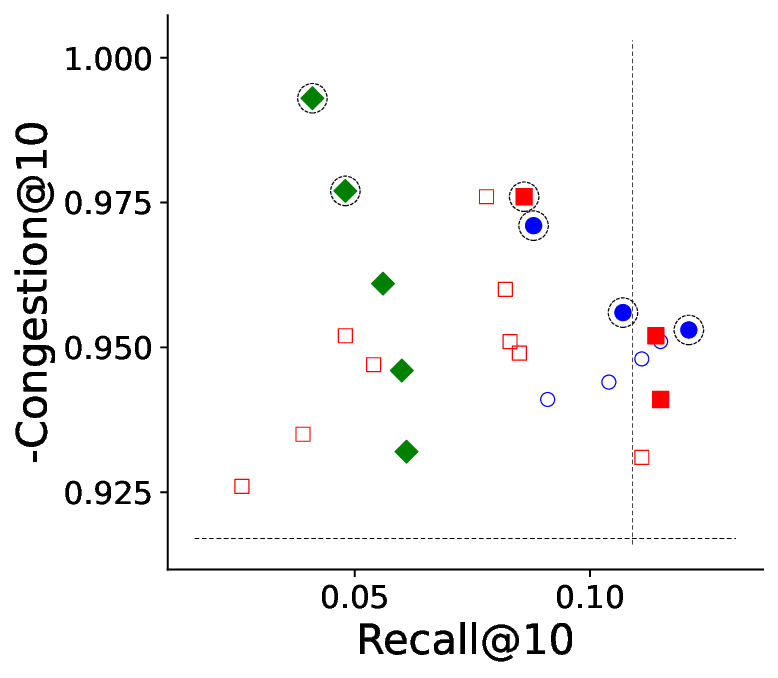}
  \end{subfigure}
  \begin{subfigure}{.33\textwidth}
  \includegraphics[width=\textwidth,height=\textheight,keepaspectratio]{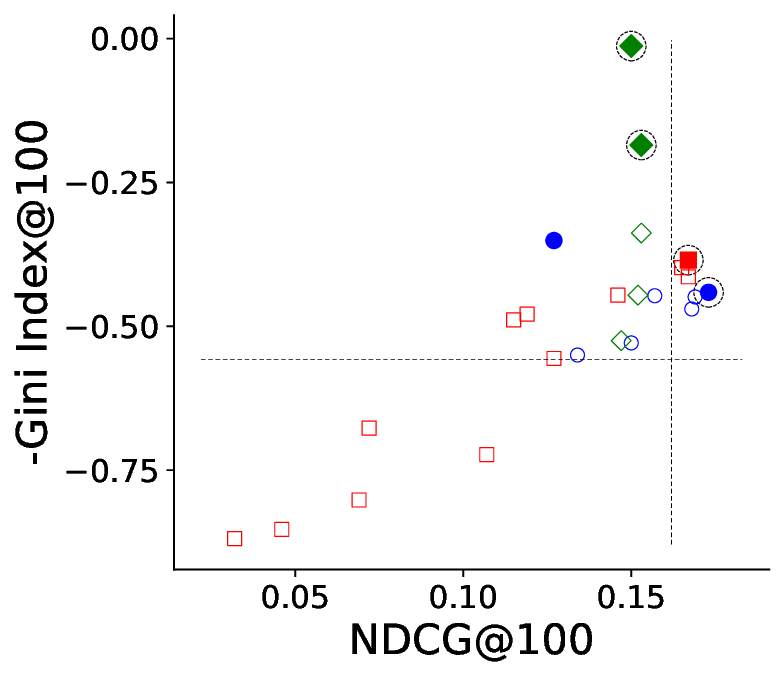}
  \end{subfigure}
    \caption{VDAB}
  \end{subfigure}

  \begin{subfigure}{1\textwidth}

  \begin{subfigure}{.33\textwidth}
  \includegraphics[width=\textwidth,height=\textheight,keepaspectratio]{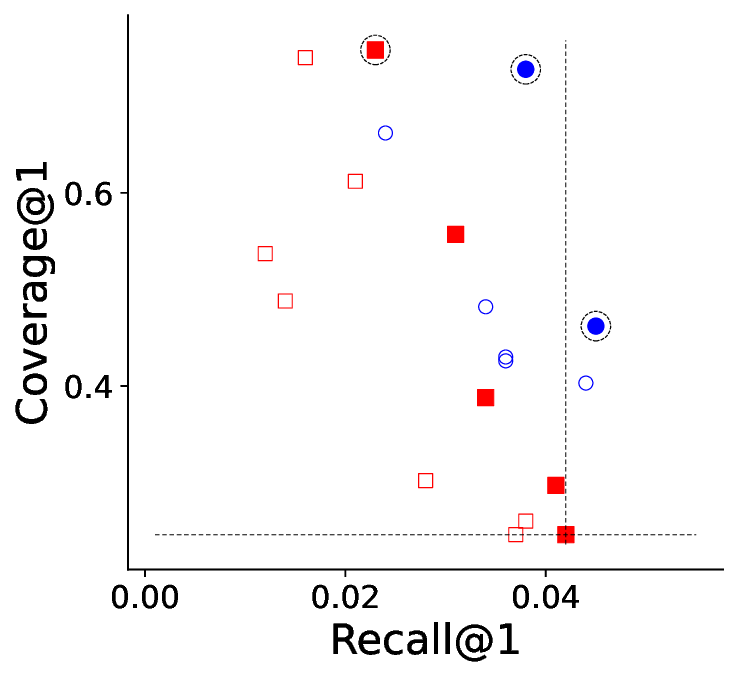}
  \end{subfigure}
  \begin{subfigure}{.33\textwidth}
  \includegraphics[width=\textwidth,height=\textheight,keepaspectratio]{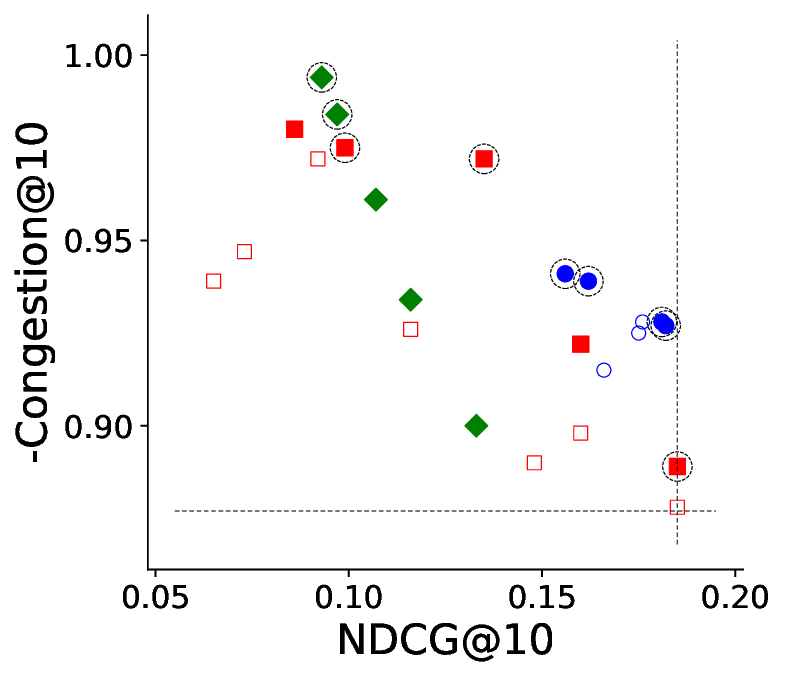}
  \end{subfigure}
  \begin{subfigure}{.33\textwidth}
  \includegraphics[width=\textwidth,height=\textheight,keepaspectratio]{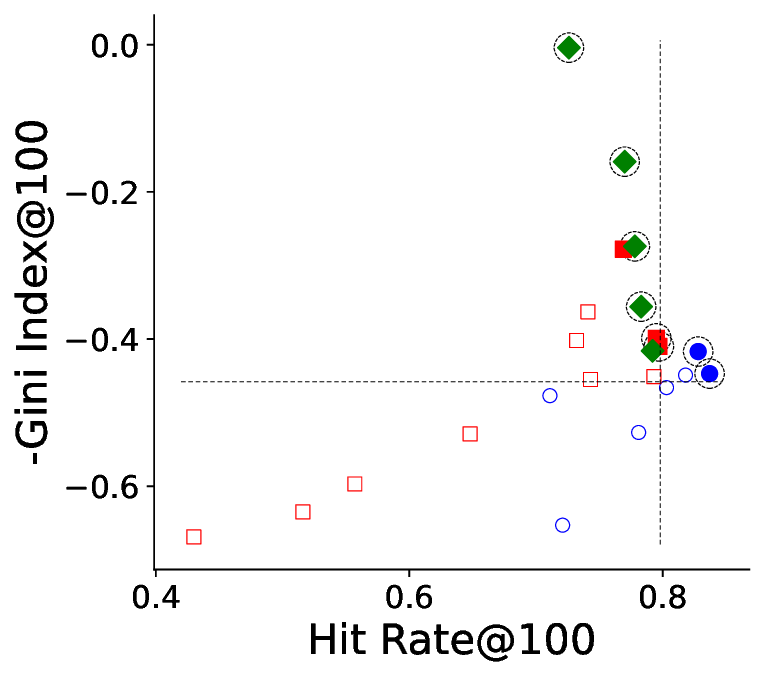}
  \end{subfigure}
  \caption{CareerBuilder}
  \end{subfigure}

  \caption{Desirability versus congestion-related measures (higher values are better). Points represent different hyper-parameter combinations. Pareto optimal points per method are solid (filled). Pareto optimal points across methods have a circle around.}\label{fig:pareto}
  \Description[Baseline comparison]{Comparing desirability measures and congestion-related measures at top-1, top-10, and top-100 job recommendations for all methods.}
\end{figure*}

Here we compare the methods in terms of the desirability measures and congestion-related measures (\textbf{Q1}).
Figure~\ref{fig:pareto} shows desirability measures against congestion-related measures for all datasets. \texttt{ReCon} mostly improves congestion-related measures, while keeping good performance (or improving over CNE in some cases) of desirability measures for some selections of hyper-parameters. We can also observe that \texttt{ReCon} is Pareto optimal for some selections of hyper-parameters. While achieving almost the same or better desirability measure performance compared to CNE, \texttt{ReCon} can result in better congestion-related measures compared to plain CNE and the baselines. 

The same conclusion could generally be established for all the desirability measures against all congestion-related measures, where for some selections of hyper-parameters, \texttt{ReCon} usually finds a good trade-off between both measures. More figures for different combinations of measures are available in the \href{https://github.com/aida-ugent/ReCon/}{online supplementary material}\footnote{\url{https://github.com/aida-ugent/ReCon/}}.

Compared to the baselines, \texttt{ReCon} has an additional advantage in that it trains a recommendation model, whereas the baselines simply generate a recommendation list by post-processing the original recommendations from a model. This makes \texttt{ReCon} particularly advantageous in real-world applications where the model requires incremental updates to account for new interactions.


\subsubsection{Execution time comparison\label{exp:execution_time}}
In this experiment, we compare the methods in terms of the execution time (\textbf{Q2}). 

\begin{figure}[t]
  \centering
  \includegraphics[width=1\linewidth,height=\textheight,keepaspectratio]{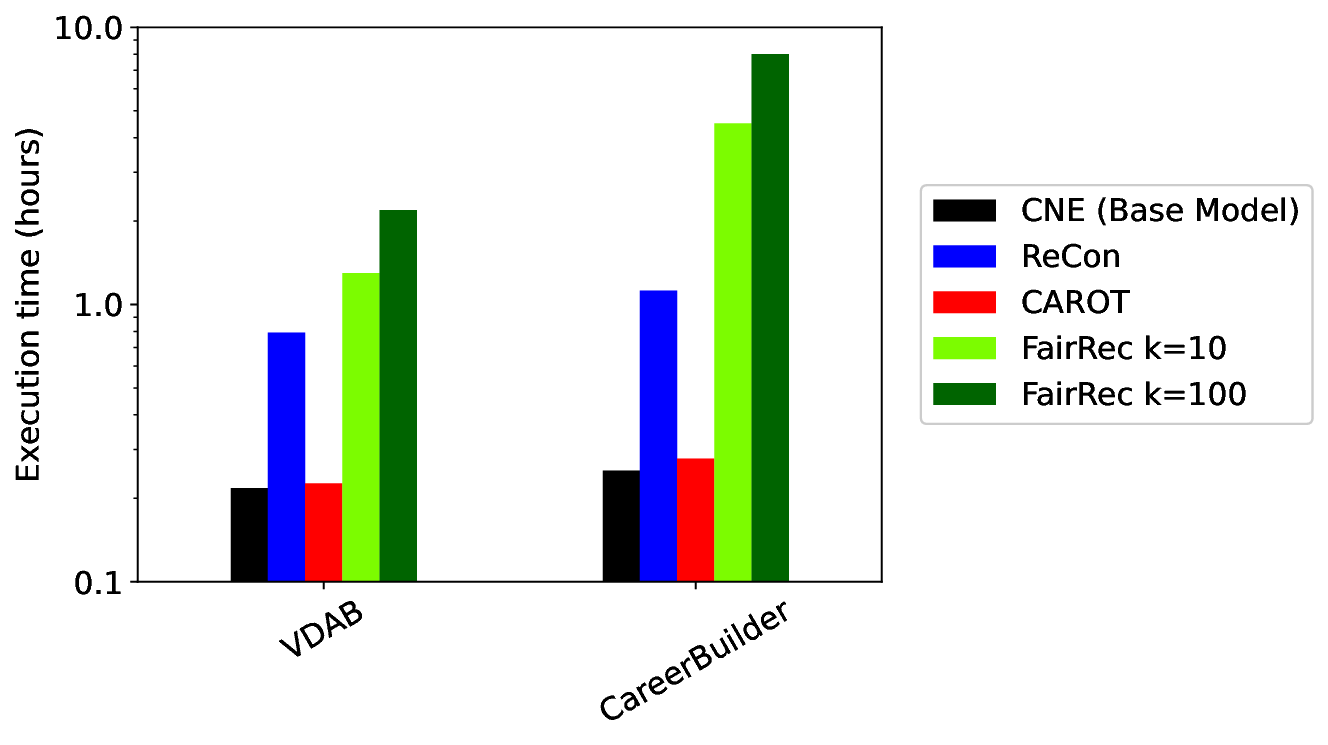}
  \caption{Execution time (log-scale) for both datasets. The execution time of each baseline is averaged over different selections of hyper-parameters.}\label{fig:exp_runtime}
  \Description[Execution time comparison]{Comparing the execution time of the methods.}
\end{figure}

Figure~\ref{fig:exp_runtime} shows the execution time in hours for both datasets. The execution time of each baseline is averaged over different selections of hyper-parameters. The training time of CNE is added to the execution time of CAROT and FairRec, as they receive the matching scores from the base recommendation model.
\texttt{ReCon} is slower than CNE and CAROT since it optimizes the optimal transport cost and the base recommendation model jointly. However, \texttt{ReCon} is faster than FairRec. 
%

\section{Related Work}\label{sec:related_work}

One important aspect of job recommendation models is that there is competition between job seekers/jobs for the same jobs/job seekers \cite{mashayekhi2022challenge}. This competition amplifies the problem of congestion with recommendations in the job market.


In \cite{borisyuk2017lijar}, a job application redistribution system was proposed for LinkedIn to prevent job postings from receiving an excessive or insufficient number of applications. To achieve this goal, the job recommendation scores were adjusted using a dynamic forecasting model that penalized or boosted scores based on the predicted number of applications. Chen et al. \cite{chen2019prediction} use a decentralized economic model to learn the scoring functions and use an optimal transport approach for the final recommendations to reduce congestion in the recommendations.

To reduce congestion in job recommendation, two-step approaches (including CAROT) are proposed in \cite{bied2021congestion,naya2021designing}. The first step uses a model to predict the recommendation scores, and the second step employs optimal transport to better distribute jobs between job seekers and generate the final recommendations. In contrast, \texttt{ReCon} integrates the congestion reduction with recommendation into a single multi-objective task.

From a fairness perspective, two recent approaches (including FairRec) \cite{patro2020fairrec,biswas2021toward} develop greedy algorithms to ensure Envy-Free job recommendations for job seekers and to guarantee a minimum exposure for job vacancies.

\section{Conclusion and Future Work}\label{sec:conclusion_future_work}
In this paper, we proposed a novel approach, \texttt{ReCon}, for reducing congestion in job recommendation systems using optimal transport theory. Our approach aimed to ensure a more equal spread of jobs over job seekers by combining optimal transport and a given job recommendation model in a multi-objective optimization problem. The evaluation results showed that our approach had good performance on both congestion-related and desirability measures. The proposed method can be a promising solution to the problem of congestion in job recommendation systems, and can potentially lead to more efficient and effective job matching.

For future work, we intend to evaluate \texttt{ReCon} using various recommendation models. Moreover, we aim to scale \texttt{ReCon} to be applicable to large-scale job market datasets by optimizing optimal transport in each batch of the training data, instead of the whole dataset. Additionally, this could reduce the training time as well.

\begin{acks}
The research leading to these results has received funding from the European Research Council under the European Union's Seventh Framework Programme (FP7/2007-2013) (ERC Grant Agreement no. 615517), and under the European Union’s Horizon 2020 research and innovation programme (ERC Grant Agreement no. 963924), from the Flemish Government under the ``Onderzoeksprogramma Artifici\"ele Intelligentie (AI) Vlaanderen'' programme, and from the FWO (project no. G091017N, G0F9816N, 3G042220). Part of the experiments were conducted on pseudonimized HR data generously provided by VDAB.
\end{acks}

\bibliographystyle{ACM-Reference-Format}
\bibliography{main_arxiv}

\end{document}